\documentclass[asus]{JAC2003}
\usepackage{graphicx}

\begin{document}

\title{STUDY OF NEAR-FIELD VIBRATION SOURCES FOR THE NLC LINAC COMPONENTS\\[-5mm]}
\author{F.~Asiri, F.~Le~Pimpec, A.~Seryi ~~SLAC, CA USA
\thanks{Work supported by the U.S. Department of Energy, Contract
  DE-AC03-76SF00515.}
 }
\maketitle

\begin{abstract}

The vibration stability requirements for the Next Linear Collider
(NLC) are far more stringent than for the previous generation of
Colliders. To meet these goals, it is imperative that the effects
of vibration on NLC Linac components from near-field sources (e.g.
compressors, high vacuum equipment, klystrons, modulators, pumps,
fans, etc) be well understood. The civil construction method,
whether cut-and-cover or parallel bored tunnels, can determine the
proximity and possible isolation of noise sources. This paper
presents a brief summary and analysis of recently completed and
planned studies for characterization of near-field vibration
sources under either construction method. The results of in-situ
vibration measurements will also be included.

\end{abstract}

\vspace{-0.2cm}

\section{Introduction}

\vspace{-0.2cm}

\begin{figure}[tbph]
\begin{center}
\vspace{-0.3cm}
\includegraphics[width=0.5\textwidth,clip=]{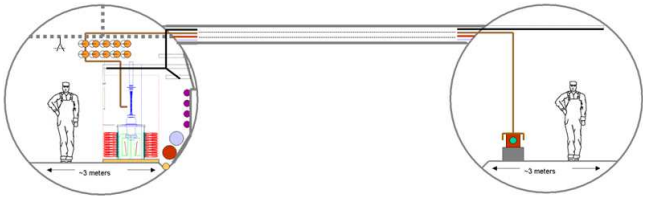}
\end{center}
\vspace{-0.6cm} \caption{Cross-section of support and beam
tunnels.} \vspace{-0.2cm} \label{2tunnel}
\end{figure}

To maintain the desired luminosity of the NLC, the focusing
components on the main Linac must be kept at a few nanometers above
a few Hz. These components can be affected by far-field (natural)
and near-field (man-made) vibration sources. This paper is
concerned only with near-field sources (e.g. mechanical and
electrical equipment, RF generating equipment, etc).  These
sources are mainly located either in the Support tunnel,
Fig.\ref{2tunnel}, or far away ($>$ 100~m) from the Beam tunnel.
The characterization of near-field vibration sources and its
effects on the main Linac components is part of an ongoing R\&D
program at NLC that is presented in this paper. The first part of
the paper will present the influence of the vibration induced by
RF power generating elements and by the RF itself. The second part
will deal with the transfer of vibration from surface to the tunnel invert.

\vspace{-0.2cm}

\section{RF Vibration Characterization}
\label{RFklystron}

At first, we will focus on the vibration contribution of high
power generating RF components. The study was carried out near the
supporting structure of the 8 pack project \cite{8packproject}.
The 8 pack project is the test bench of klystrons and their
modulator to produce high power RF for the Next Linear Collider
Test Accelerator (NLCTA). One of the geophone sensors was placed
at the base of the modulator at 76~cm above the concrete floor and
the second one was located on the concrete floor at $\sim$1.2~m
away from the base of the modulator support, or at its base (0~m).
The signals of the Mark-4 geophones were measured simultaneously.

We performed tests in two different conditions of the modulator.
In the first case, the water cooled modulator was under a 20~kV
voltage and was running at 30~Hz or 10~Hz (at this voltage the
klystrons was not conducting, thus did not deliver RF power). In
the second case the modulator was running at 10~Hz and was under
400~kV (the klystron was delivering $\sim$40~MW of RF power at
1.6~$\mu$s).

\begin{figure}[tbph]
\begin{center}
\vspace{-0.3cm}
\includegraphics[clip=,totalheight=5.7cm]{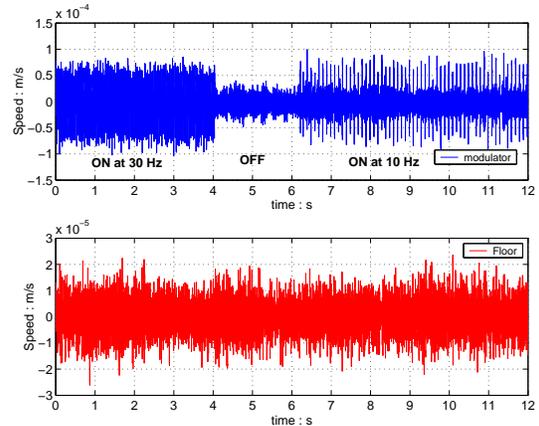}
\end{center}
\vspace{-0.6cm} \caption{Response of the geophones when the
modulator was running at different frequencies or was switched
off. Note the difference of the scales.} \vspace{-0.2cm}
\label{speedtimesensor}
\end{figure}

Fig.\ref{speedtimesensor} shows the response of the sensors vs
times for the modulator under a 20~kV. In this case the modulator
was running at different frequencies. The seismometer located at
the base of the modulator clearly shows vibrations due to
modulator pulsed operation, while the geophone located on the
concrete floor does not indicate any change due to the modulator
running conditions.

\begin{figure}[tbph]
\begin{center}
\vspace{-0.3cm}
\includegraphics[clip=,totalheight=5.9cm]{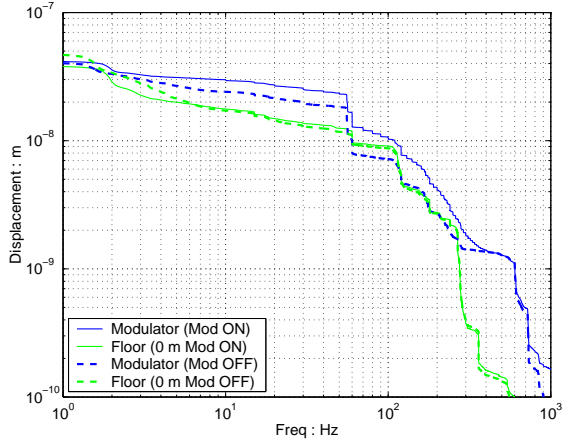}
\end{center}
\vspace{-0.6cm} \caption{Average Integrated Displacement (AID) of
the seismometers when located at the base of the modulator and on
the floor at $\sim$1.2~m of the supporting structure. The
modulator is under 20~kV.} \vspace{-0.2cm} \label{Displc8packjan}
\end{figure}

\begin{figure}[tbph]
\begin{center}
\vspace{-0.3cm}
\includegraphics[clip=,totalheight=5.9cm]{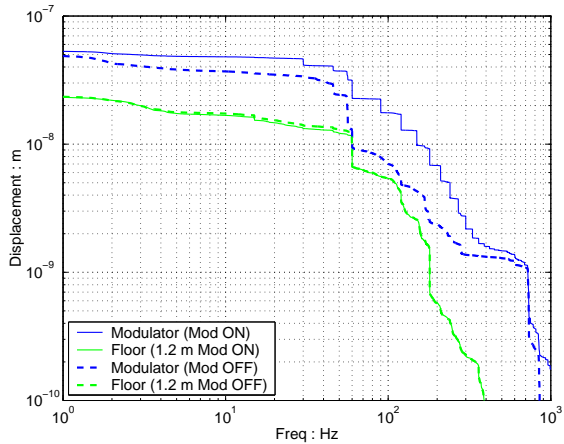}
\end{center}
\vspace{-0.6cm} \caption{AID of the seismometers when located at
the base of the modulator and on the floor 0~m of the supporting
structure. The modulator is under 400~kV.} \vspace{-0.2cm}
\label{Displc8packfeb}
\end{figure}

Fourier analysis also does not reveal any additional noises due to
modulator on the floor.  Fig.\ref{Displc8packjan} and
Fig.\ref{Displc8packfeb} show the average integrated displacement
(AID) of the two seismometers. The blue lines are the response of
the sensor placed at the base of the modulator, where the
difference between the modulator on and off cases is clearly seen.
The green lines are the response of the sensor placed on the
concrete floor either at $\sim$1.2~m or at the base of the stand
(0~m), and the on and off cases overlap.

Interesting that, when the modulator is under 400~kV and
delivering power to the klystron, its vibration is slightly less
than for a 20~kV running modulator. One can also note that
background noise was different in the conditions of
Fig.\ref{Displc8packjan} and Fig.\ref{Displc8packfeb} (see green
curves below 60~Hz) which was due to different level of activity
of the construction crew working at NLCTA.

The main conclusion of the modulator vibration study is that the
transmission of vibration from the modulator to the concrete floor
is not significant, and, on the level of background noise at
NLCTA, not noticeable.

We also studied vibrations of accelerating structures due to
cooling water (reported earlier, see \cite{lepimpec:nanobeam02}),
and due to RF pulse (presented below). When the klystrons are
delivering power, the RF heating produces acoustical vibration in
the accelerating structure. Possible vibration produced by the RF
were measured by 3 piezo-accelerometers placed on the accelerating
travelling wave structure bolted on its support or girder, placed
on the girder, and placed on the waveguide. The X-rays coming out
from the structure when filled by the RF did not affect the
measurements of our accelerometers.

\begin{figure}[tbph]
\begin{center}
\vspace{-0.003cm}
\includegraphics[clip=,totalheight=5.9cm]{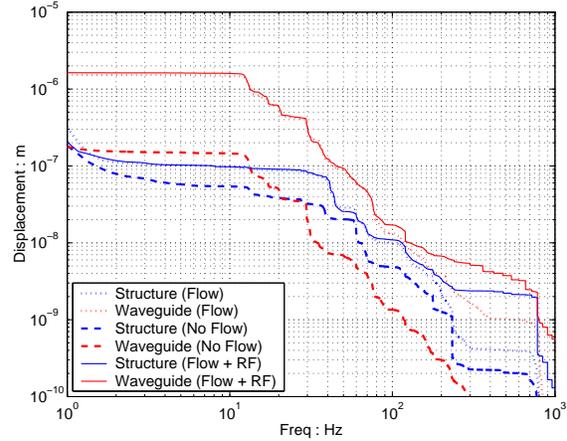}
\end{center}
\vspace{-0.6cm} \caption{AID for the structure and its connected
waveguide when cooled or not with water (dot and dashed line) then
when cooled and fed with RF.} \vspace{-0.2cm} \label{RFinNLCTA}
\end{figure}

These measurements, Fig.\ref{RFinNLCTA}, show that feeding a
structure (60~cm long travelling wave structure H60Vg3R), via its
waveguide, with 100~MW at 400~ns of RF power at 60~Hz
(corresponded to about 70~MV/m accelerating gradient) does not
lead to any significant increase of vibration in comparison to
vibrations produced by cooling water. The water induced vibrations
dominate, but they are tolerable, since, with an appropriate
design, the vibration transmission to the linac quadrupoles is
rather small \cite{lepimpec:nanobeam02}. Vibration of the loosely
supported RF waveguide is higher than of the RF structure, but
apparently this does not significantly increase vibration of
either structure or the quadrupoles.

\section{Vibration transfer from surface to beam tunnel}

One of the questions to be addressed is whether there is a
significant difference in the vibration attenuation
characteristics between a tunnel bored at great depth (in bedrock)
or excavated in cut-and-cover construction at lesser depth, with
regard to vibrations generated at the surface. A study was carried
out at SLAC \cite{colingordon}, representing the vibration
attenuation characteristics of the soil, assuming the SLAC beam
line housing to be representative of a cut-and-cover construction.
The second vibration measurement study is currently underway in
the Red Line tunnels in Los Angeles. This study will establish the vibration
characteristics between two tunnels, along the tunnel as well as
from surface to the tunnel. The following is a brief presentation
of data from the study carried out at SLAC.

\begin{figure}[tbph]
\begin{center}
\vspace{-0.3cm}
\includegraphics[width=0.5\textwidth,clip=]{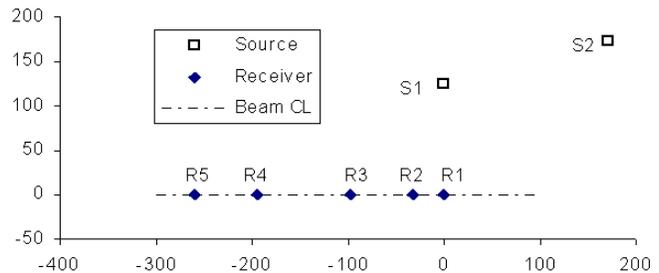}
\end{center}
\vspace{-0.6cm} \caption{Relative locations of sources and
receptors (scale in feet).} \vspace{-0.2cm} \label{relativeloc}
\end{figure}

Fig.\ref{relativeloc} shows the relative locations of sources and
the receivers. The receivers locations were on the floor along the
centerline of the beam. The sources were on the ground surface.
Drive location S1 was at the same approximate elevation as of the
Klystron gallery floor, which lies about 36 ft (~11 m) above the
elevation of the receiver locations. Drive location S2 was about
10 ft (~3 m) uphill from S1.

\begin{figure}[htbp]
\begin{center}
\includegraphics[width=0.5\textwidth,clip=]{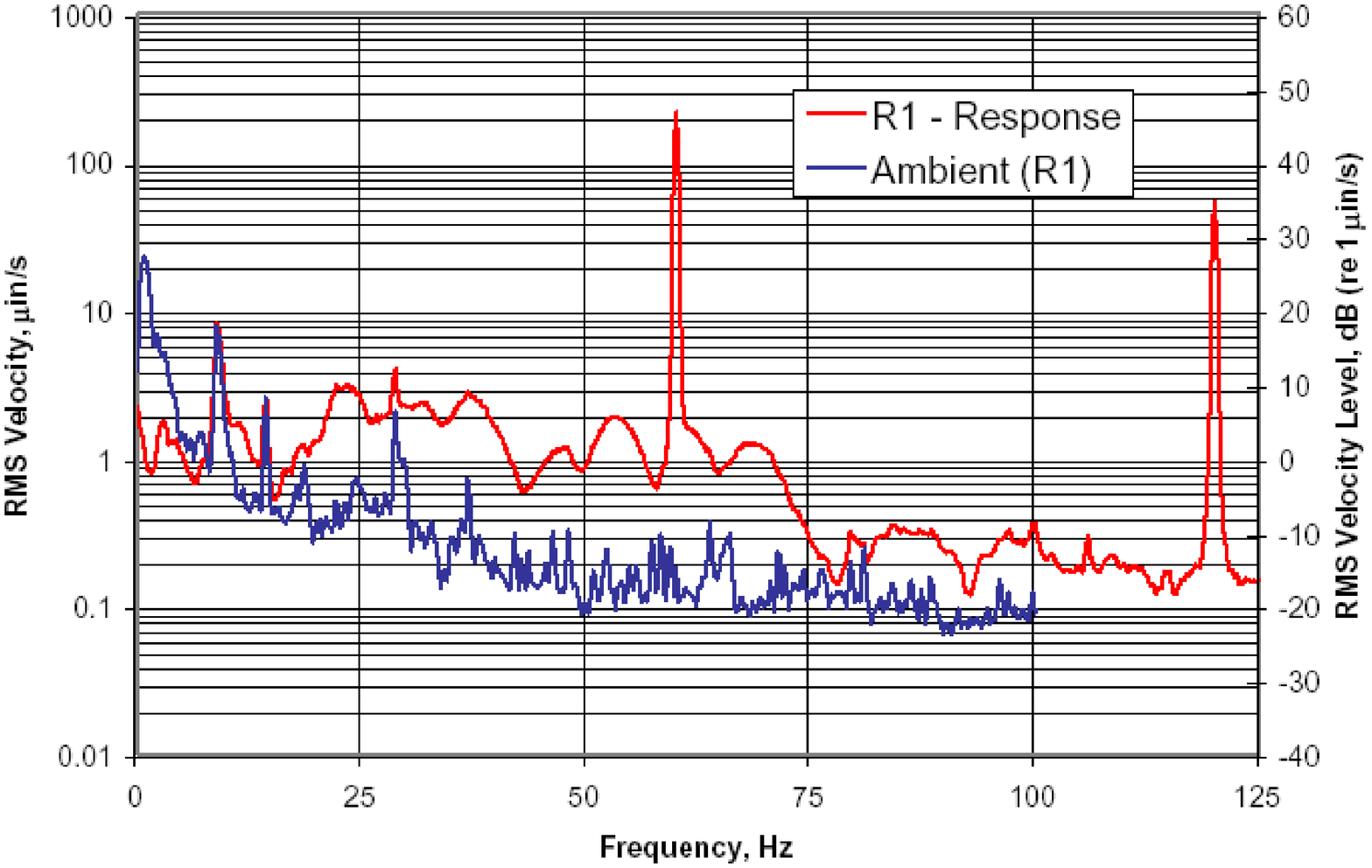}
\end{center}
\vspace{-0.5cm} \caption{Typical response at R1 to hammer blow at
S1, with ambient at R1.} \vspace{-0.2cm} \label{Hammerblow}
\end{figure}

Fig.\ref{Hammerblow} shows two spectra. One is the ambient
measured at the receiver location R1. The other is a fast Fourier
transform of the response to several hammer blows. Several
observations may be made. The peaks at 60~Hz and 120~Hz are
electronic artefacts. The peaks centered at $\sim$8.75~Hz and at
$\sim$14.25~Hz in both ambient and response have a nearly
identical amplitudes. The peak at 18.75~Hz has a slightly
different amplitude. At these frequencies, the ``response'' (red
line) is being governed more by ambient than by the input force,
so the transfer functions derived from the ``response'' will be
invalid. The ``ambient'' (blue line) at frequencies less than 7 Hz
(in this case) lies above the ``response''. This suggests that
there is some variability to the ambient environment at these low
frequencies, and these will degrade the accuracy of transmission
function at lower frequencies.

\begin{figure}[htbp]
\begin{center}
\includegraphics[width=0.45\textwidth,totalheight=6.5cm,clip=]{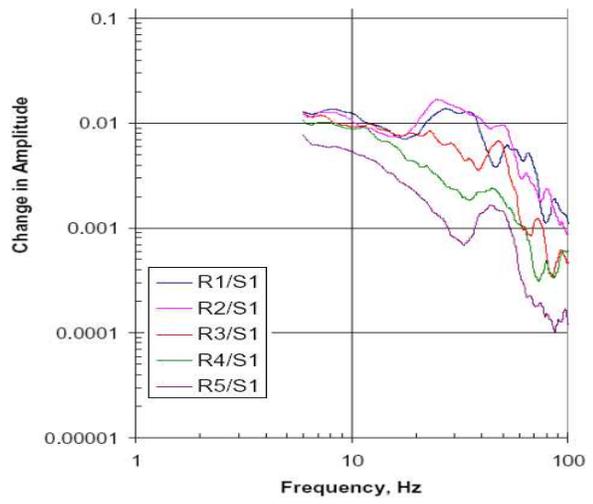}
\end{center}
\vspace{-0.5cm} \caption{Log-mean ground transmission from the
drive point S1 -- heavily smoothed.} \vspace{-0.002cm}
\label{GroundTransmis}
\end{figure}

Fig.\ref{GroundTransmis} shows a smoothed spectra which were
calculated by log averaging the amplitudes over a 10~Hz interval
centered on the plotted point. It provides the transfer function
(showing change of amplitude) measured at R1 to R5 using S1 as a
drive point, Fig.\ref{relativeloc}.

\begin{table}[htbp]
\begin{tabular}{|c|c|c|c|c|}
\hline {Source (S1) to} & \multicolumn{4}{|p{155pt}|}{Attenuation
at Given Frequency} \\ \cline{2-5} {Receiver(m)} & 10 Hz& 20 Hz&
30 Hz&
60 Hz \\
\hline R1(39)& 0.024& 0.0084& 0.012&
0.005 \\
\hline R2(40.2)& 0.012& 0.012& 0.014&
0.004 \\
\hline R3(48.6)& 0.011& 0.0084& 0.006&
0.001 \\
\hline R4(69.9)& 0.010& 0.004& 0.002&
0.001 \\
\hline R5(86.7)& 0.005& 0.002& 0.0009&
0.0003 \\
\hline
\end{tabular}
\caption {Attenuation factor, refering to Fig.\ref{relativeloc}.}
\label{tblea}
\end{table}

The Table~\ref{tblea} summarizes the apparent attenuation at
several frequencies at which mechanical equipment usually
operates.  The figures in Table~\ref{tblea} represent the
attenuation factor A for a vibration with its source near S1
propagating along the same path. Let us give an example how such
data can be used. Suppose a pump is installed at S1, and it
produces vibrations at 30~Hz of amplitude X. The amplitude at
30~Hz that we would measure at R5 would be the greater of either
ambient or 0.0009X. Suppose we take the ambient measured shown for
the tunnel in (Fig.\ref{Hammerblow}) as representative the tunnel
in general. The amplitude at 30~Hz is about 1.5~$\mu$m/s. If we
were to place a pump at S1 and be sure to avoid having its
vibrations exceed ambient at R5, we would need to impose a limit
on the resulting vibration at S1 of 1.5/0.0009=1.7 mm/s.

\section{Conclusion}

It has been shown that the vibration transmitted by the RF
generating equipment to the floor is insignificant. Hence,
klystrons and or modulators running in the Support tunnel of the
NLC should not effect alignment of the Linac. Vibration
contribution of an RF pulse to an accelerating structure has also
been found negligible relative to water-cooling. Thus, it leaves
electrical and mechanical rotating equipment as possibly a
dominating source of vibration. The attenuation factors presented
in the paper can be used in planning stage of the NLC project for
specifying and locating the mechanical rotating equipment, as well
as to assess their vibration effects on the focusing components on
the Main Linac and provide a means for establishing the vibration
budgeting scheme for the project.

\end{document}